\documentclass{epl}

\usepackage{graphicx}
\usepackage{amsmath}
\usepackage{amssymb}

\title{Realistic time correlations in sandpiles} 
\author{Marco Baiesi \and Christian Maes}
\institute{Instituut voor Theoretische Fysica, K.U.Leuven,
Celestijnenlaan 200D, B-3001 Leuven, Belgium
}

\begin{document}

\maketitle

\begin{abstract}
A ``sandpile'' cellular automaton achieves complex temporal
correlations, like a $1/f$ spectrum, if the position where it is perturbed 
diffuses slowly rather than changing completely at random, showing that
the spatial correlations of the driving are deeply related to the 
intermittent activity.
Hence, recent arguments excluding the relevance of self-organized criticality 
in seismicity and in other contexts are inaccurate.
As a toy model of single fault evolution, and despite of 
its simplicity,
our automaton uniquely  reproduces the scaling
form of the broad distributions of waiting times between earthquakes. 
\end{abstract}

Many complex natural
phenomena exhibit quiet periods with a slow loading of the system
separated by events in which a rapid evolution takes
place~\cite{bak_book,sornette_book}. Some examples are
earthquakes~\cite{bak02:_unified_old,corral03:_unified}, solar
flares~\cite{BPS_05}, creep experiments
with cellular glasses~\cite{maes98:_glass} and even with  
piles of rice~\cite{aegerter03:_rice-exper}, and rain 
falls~\cite{peters_rain}. 
In these and in other natural and social systems one 
finds that the intensity of the events and the spatial and
temporal scales separating them may vary over broad ranges and 
often are scale-invariant. 
The similar phenomenologies and the nonequilibrium nature
 suggest that a common mechanism is emerging in
different systems. An interesting candidate is
self-organized criticality
(SOC)~\cite{bak_book,sornette_book,btw}, a theory essentially focussing on the 
occurrence of fast avalanches of nonlinearities, which have been well 
visualized in simple cellular automata (CA) called ``sandpiles''. 

CA and their probabilistic versions are model-systems
that have been used quite extensively in the field of complex 
systems~\cite{gutowitz91,frisch86}.
Good reasons include the possibility of reliable simulations and the
conceptual simplicity of the dynamical rules.  Furthermore, 
one has strong
indications that the microscopic details of nature are not
all-important in deciding certain macroscopic features.  Symmetry
properties, conservation laws and considerations on the right scale of
description matter much more.  CA incorporate these
essential ingredients and express in the most strongest form we know
today how emergent behavior can be very rich, varied and complex
even if based on few simple dynamical rules on a discrete architecture.

A feature distinguishing  several sandpile models from other CA
is the scale-free distribution of their avalanche sizes. 
This resembles the distribution of energy released by earthquakes, for
example.
On the other hand, temporal
correlations between avalanches have been far less investigated, until
Boffetta {\em et al.}~noticed that
realistic correlations are not found in time series of 
some sandpiles~\cite{boffetta99}.
Indeed, their avalanches have interoccurrence times distributed exponentially,
and the time series appears as a Poisson process of uncorrelated events. 
This fact was used to argue that SOC cannot be an interpretation
of solar flares activity~\cite{boffetta99} and, more recently, that SOC is
not related to earthquakes~\cite{yang2004}.

After the critique to SOC by Boffetta {\em et al.}~\cite{boffetta99} 
there has been a wave of 
investigations on the temporal properties of sandpiles and other
SOC models. 
Nowadays we know that there
are models with interesting temporal correlations, like the
$1/f$ decay in power 
spectra~\cite{zhang00:_1f,davidsen-paczuski02:_1f,woodard05:_building},
or like power-law tails in distributions of waiting times between
events~\cite{norman01:_model_waiting,sanchez02:_wait-SOC,fragos04:_model_Ellerman,hedges05:_OFC_waiting,lippiello05:_SOC_memory,maya05}, 
foreshocks and aftershocks like
for earthquakes~\cite{hergarten02:_OFC_aftershock} or even
features typical of turbulence~\cite{btw_turb,fabio}.

Correlated drivings can give rise to non-exponential waiting time 
distributions between events~\cite{sanchez02:_wait-SOC}, which 
e.g.~appeared in models of solar activity. In one case the strength of
the driving perturbation in the Lu-Hamilton model was slowly
varied at random~\cite{norman01:_model_waiting}. In another model
of ``Ellerman bombs''~\cite{fragos04:_model_Ellerman}, directed
percolation was one mechanism increasing the number of driving
points. These examples further confirm and illustrate the
coexistence of criticality and complex time correlations
but it is likely that the correlations intrinsic in these driving mechanisms
are simply inherited and show up directly in the output of the automaton.

In this Letter we stress that a completely
random driving of sandpiles is not expected to be a realistic
feature but rather an artificial feature put by hand. In other words,
random drivings can be an arbitrary, {\em a priori} limitation of the 
automata. 
Usually sandpiles are driven either at completely random positions
or from a fixed single site, but none of these cases reflects the
forms of slow loading in real systems, like the crust of the Earth
or the solar corona. 
In seismicity, for example, one observes a rich scaling 
picture for the waiting time 
distributions~\cite{bak02:_unified_old,corral03:_unified},
involving location, temporal occurrence, and magnitude of events.
In order to have satisfactory CA models of earthquakes,
it is therefore desirable to remove excessive randomness from the 
properties of their driving.

Below, we show how extremely simple spatial correlations in the input
drive can generate a complex time series as an output. 
Thus, the focus is  on the {\em spatial} properties of the nucleating
point of the avalanches. 
As far as we know, we
provide the first example of $1/f$ noise in the avalanche activity
of a non-running, classical sandpile. Furthermore, we observe
broad distributions of waiting times between events also of small sizes. 
Most importantly,
these distributions display a scaling behavior with intensity 
thresholds, analogous to those recently
found for earthquakes~\cite{corral03:_unified} and for solar 
flares~\cite{BPS_05}. This feature makes at present this new model unique,
and corroborates its interpretation in geophysical terms.

The dynamics of the model strongly  resembles the one of the sandpile 
originally discussed by Kadanoff {\em et al.}~\cite{kadanoff89}: 
each site $i$ of a linear lattice of
side $L$ holds an integer number $h_i$. In the sandpile jargon, $h_i$ is
normally called ``number of grains'', but
this distribution as well may be thought as the potential energy profile in 
an object mainly extended in one dimension, like a fault. 
Stable
configurations fulfill a local stability condition on the discrete
gradient between every pair $(i, j)$ of nearest neighbors,
\begin{equation}
h_i - h_j < H \label{eq:inst}
\end{equation}
 where constant $H$ is a
threshold. A local instability not fulfilling~(\ref{eq:inst}) is
resolved by a toppling, which consists in moving $\alpha$ grains
from $i$ to $j$, i.e., $h_i\to h_i - \alpha$ and $h_j \to h_j +
\alpha$. In the geophysical interpretation,
the toppling thus models a redistribution of energy that takes
place after some local elastic threshold within a fault is overtaken.

Stress increase appears as a local increase of potential energy during
time step $t$ at position $i'$, causing $h_{i'}(t) = h_{i'}(t-1)+1$. This
eventually leads to a violation of~(\ref{eq:inst}) between site
$i'$ and one or more of its nearest neighbors $j$. Thus, a toppling
takes place between each unstable pair $(i',j)$, and $j$ in turn
may also become unstable. One then iterates this procedure by making
all topplings in parallel as long as some pairs of sites
violating~(\ref{eq:inst}) are present~\cite{note0}. The whole set of
updates giving rise to a new stable configuration represents an
avalanche (earthquake) at time $t$, with an area $a$ equal to the number of
sites that toppled at least once and a size $s$ (energy released)
equal to the total number of toppling. As in the standard sandpile 
scenario, the long
time scales of the driving are thus completely separated from the
ones of the avalanches. This is the appropriate limit for models of
earthquakes, while 
mixing of time scales is present 
in solar activity~\cite{BPS_05}, 
and other models of SOC are possible~\cite{maya05}.

\begin{figure}[!bt]
\begin{center}
\includegraphics[angle=0,width=5.5cm]{fig_EPL_1a.eps}
\hskip 0.3truecm
\includegraphics[angle=0,width=5.5cm]{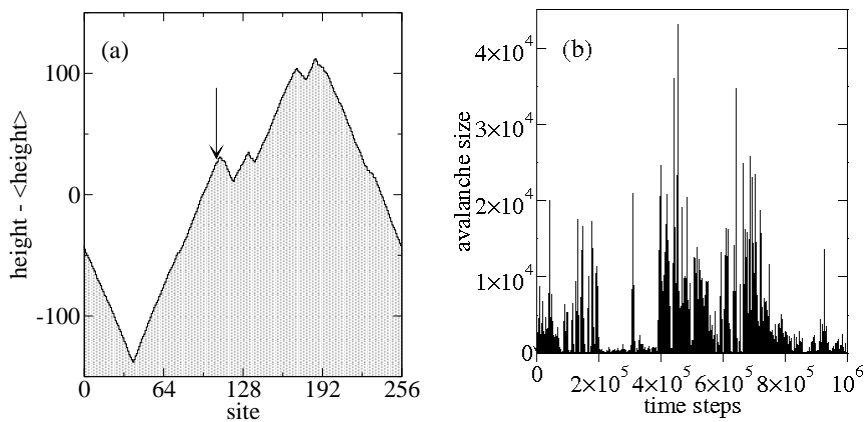}
\caption{(a) Snapshot of a sandpile profile ($L=256$) around its
average height. The arrow points to the position $i'$ where grain
addition was taking place. (b) Example of time series of the avalanche sizes
for a sandpile with $L=2048$. \label{fig:1}}
\end{center}
\end{figure}

A novel feature distinguishing the model proposed here from previous ones is
that the site where a grain is added coincides with
the position of a random walk: $i'(t)$  with equal probability is
drawn from one of the nearest neighbors of $i'(t-1)$. This feature again has a 
counterpart in seismicity, where it is known that epicenters are spatially
clustered and that aftershocks slowly diffuse after large 
events~\cite{scholz_book}.

The particular driving introduced here
makes the model critical also with periodic boundary conditions
(BC)~\cite{note_BC}. 
We adopted periodic BC because there cannot be effects related 
to a fluctuating
distance between a predefined boundary and the point where grains are
added. For our
purposes it is enough to show results on the 
sandpile in one dimension, with $H=4$ and $\alpha=2$.
According to our data, the versions with
different $H$ or $\alpha$ give similar results.
The same remark applies to choices of different mobility; as long
as the walker diffuses slowly,
no change occurs in the basic aspects of what follows.

A typical sandpile profile is plotted in Fig.~\ref{fig:1}(a).
Configurations like that are reached after an appropriate period
of fueling of an empty lattice. That period was then not
considered in the following statistical analysis. The sandpile
profiles observed with periodic BC usually are an alternated
 sequence of uphills and downhills, thus alternating local minima and maxima.
Each patch with the same trend is like a profile of a classical
sandpile with open BC, which oscillates around a typical slope.
However, grain addition and avalanches make
patches to dynamically evolve, merging them or
splitting them into shorter pieces.

\begin{figure}[!bt]
\begin{center}
\includegraphics[angle=0,width=7.7cm]{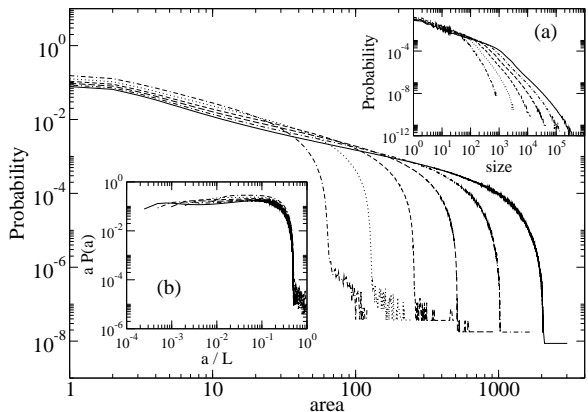}
\caption{Probibility density of the area of avalanches, for lattices with $L=256$,
$512$, $\ldots$, $4096$, from left to right.
The small secondary tail of $P(a)$ is coming from
rare avalanches with $a>L/2$ involving grains falling onto two sides of a
mountain.
Insets: (a) 
Size distributions for the same set of $L$ (curves are smoothed).
(b) Collapse of $P_L(a)$, according to Eq.~(\ref{eq:Pa}) with $\tau=D=1$,
onto a  scaling function $F$.
\label{fig:Pa}}
\end{center}
\end{figure}

A time series of the avalanche sizes in the stationary regime is
shown in Fig.~\ref{fig:1}(b). It has an intermittent behavior with
bursts of activity and temporal clustering of large avalanches,
features due to time correlations, as shown below. 
The probibility density of sizes for several $L$'s, $P_L(s)$,
are shown in Inset (a) of Fig.~\ref{fig:Pa}
and display multiscaling~\cite{tebaldi99:_multisc_BTW} while 
the area probability density $P_L(a)$ (Fig.~\ref{fig:Pa}) 
obey simple finite size scaling
\begin{equation} 
P_L(a) \simeq a^{-\tau} F\left( a / L^D \right) 
\label{eq:Pa} 
\end{equation} 
as confirmed by a collapse onto a single scaling function
$F(x)$ in the plot of $a P_L(a)$ vs $a/L$ (see the Inset (b) of
Fig.~\ref{fig:Pa}). The model thus has a critical behavior, 
with $\tau \simeq D \simeq 1$~\cite{note1}.

\begin{figure}[!t]
\begin{center}
\includegraphics[angle=0,width=7.8cm]{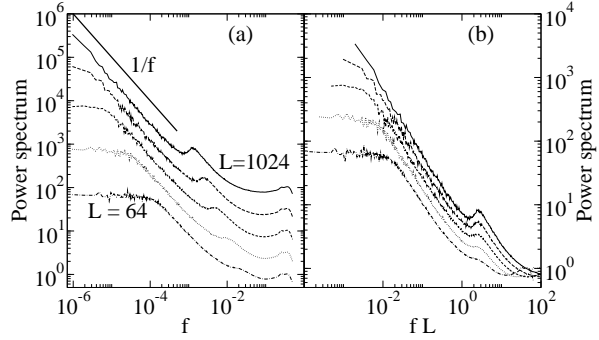}
\caption{Log-log plot of the power spectrum of the time series of avalanche
sizes, for $L=64$, $128$, $256$, $512$, and $1024$,
(a) vs the frequency and (b) vs $f L$. 
In (a) curves have been offset half decade from each
other, and a straight line indicates a $1/f$ scaling. 
For the estimates of $S(f)$, we used time windows with $2^{21}$ data.
\label{fig:Power-s}}
\end{center}
\end{figure}

Complex temporal correlations and long memory 
are revealed by the nontrivial form
of the power spectrum $S(f)$ of the time series of avalanche 
sizes $s(t)$ (with $s(t)=0$ if at time $t$ there were no 
topplings).
Fig.~\ref{fig:Power-s}(a) shows that at quite low frequencies 
$S(f) \sim 1/f^{\gamma}$ with $\gamma = 1.00(1)$. 
In Fig.~\ref{fig:Power-s}(b) we plot $S(f)$ vs $f L$ to show that the
$1/f$ region becomes broader when $L$ in increased:
it is bounded on the right by a bump at $f \sim 1/L$, 
while the crossover frequency to a flat spectrum at very low $f$ 
scales approximately as $1/L^2$. 
Almost equal spectra are observed for the time 
series of avalanche areas, confirming that
the basic origin of that $1/f$ spectrum is in the temporal correlations of
events rather than in their actual intensities. We remind that a
$1/f$ spectrum is observed in many natural phenomena~\cite{montroll84:_1f}, 
and it is regarded as one of the most complex forms of time correlation.
It is remarkable that the long-range memory in the model arises by
the non-trivial, self-organized structure encoded in the sandpile profile,
in agreement with a basic idea of SOC theory, namely that
nonequilibrium extended systems can self-organize to complex regimes
even if their dynamical rules are simple.

\begin{figure}[!t]
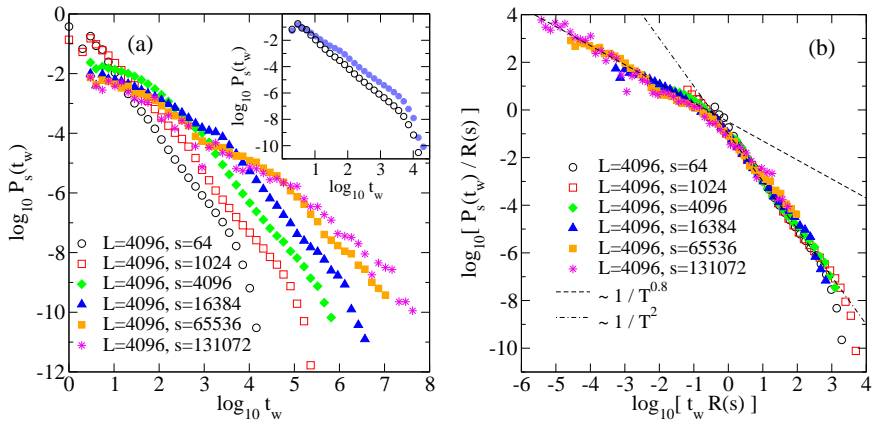

\begin{center}
\includegraphics[angle=0,width=5.7cm]{fig_EPL_4a.eps}
\hskip 2mm
\includegraphics[angle=0,width=5.5cm]{fig_EPL_4b.eps}
\caption{(a) Log-log plot of
distributions of waiting times between events with size $\ge s$,
for $L=4096$ and various $s$. Inset: the cases ($L=256$, $s=64$; $\bullet$)
and ($L=4096$, $s=64$; $\circ$)
are compared.
(b) Log-log plot of the same distributions, 
but rescaled with the rate of events $\ge s$.
\label{fig:Ptw}}
\end{center}
\end{figure}

The other form of temporal correlation concerns the waiting time
distribution between events which has a power-law tail, rather
than an exponential decay as for uncorrelated events. The waiting
time distributions $P_s(t_w)$ between avalanches of size $\ge s$ in a
sandpile with $L=4096$
are plotted in Fig.~\ref{fig:Ptw}(a) for some thresholds $s$
ranging from $2^8=64$ to $2^{17}=131072$. Already for a small
fixed $s$, the distributions have a power-law tail, which does not
disappear when larger $L$ are considered.
In the Inset of Fig.~\ref{fig:Ptw}(a), for example, one can see that
$P_{s=64}(t_w)$ has the same power-law tail for $L=256$ and $L=4096$.
Interestingly, in fact when $s\gtrsim L/2$ (steepest part of $P_L(s)$, 
see Inset (a) of Fig.~\ref{fig:Pa})
in $P_s(t_w)$ there are two regimes with two different power law decays, 
as one observes in the analysis of waiting times of
earthquakes~\cite{corral03:_unified} and for solar
flares~\cite{BPS_05}. As in these cases, we find it appropriate to
rescale times by multiplying them by rates of events $\ge s$,
denoted as $R(s)$\cite{note_Rs}.

The rescaled times $T = t_w R(s)$ appear to be the natural ones,
because $P_s(t_w) / R(s)$ vs $t_w R(s)$ collapse quite well onto a
single scaling curve $G(T)$ [see Fig.~\ref{fig:Ptw}(b)]. 
Since $G(T) dT = P_s(t_w)
d t_w$, we see that $G(T)$ is the probability density of $T$, not depending
anymore on thresholds. Hence,  a
unifying scheme can be applied to all thresholds also in our sandpile model, 
showing that
there is a basic self-similar mechanism taking place. 
For short rescaled times $G(T)\sim T^{-\beta_1}$ with $\beta_1 = 0.8(1)$,
while $G(T)\sim T^{-\beta_2}$ with  $\beta_2 = 2.0(2)$ for $T\gg 1$. 
(for earthquakes Corral found $\beta_1 = 0.9(1)$ 
and $\beta_2 = 2.2(1)$~\cite{corral03:_unified}).
The crossover between the two power-law regimes is around $T^*\approx 0.7$.
Thus, for $t_w$ much smaller than the
average occurrence time $\overline t(s) = R(s)^{-1}$ of events
with size $\ge s$ (i.e.\ $T \ll 1$), the correlations between
events are qualitatively different from the ones at $t_w \gg
\overline t(s)$.

A mechanism giving rise to strong temporal correlations between
avalanches should be their spatial overlap, since the memory of
past activity is stored in the height profile of the sandpile. By
dropping grains at a point slowly diffusing, as we do, a fresh
part of the profile is almost always found, probably enhancing the
correlation with past activity. In some classical
models driven at random points, scale-free waiting time
distributions are observed between sufficiently
large avalanches~\cite{sanchez02:_wait-SOC}, which indeed  
have a consistent probability to overlap with previous ones. 
On the other hand,
that argument does not really explain why the $1/f$ part of the
spectrum of our sandpile extends to very low frequencies.

We have mainly discussed SOC in CA, but
a different class of SOC models also supports the view that space
and time correlations are related to each other. We are referring to
models with ``extremal dynamics''. In these models each unit
carries a continuum variable and the unit closer to instability is
always the one relaxing (toppling). The Olami-Feder-Christensen
(OFC) model of earthquakes~\cite{olami92:_OFC} is 
representative of extremal dynamics. 
Whether criticality in the OFC model is only apparent or persists
 in the limit $L\to \infty$ is an open issue. 
Nevertheless, the OFC model has complex
patterns in time~\cite{hergarten02:_OFC_aftershock,hedges05:_OFC_waiting}  
and space: indeed, by its
nature, the OFC map self-organizes also the position of the
epicenters (driving points), which form sequences deeply linked
to past activity~\cite{peixoto04:_epic_OFC}. Similarly, in the
Bak-Sneppen model of evolution, distances between extremal sites
follow a Levy flight, and $1/f^{\gamma}$ spectra are
found with $0<\gamma<1$~\cite{paczuski96:_bigPRE}.
Recently, Lippiello {\em et al.}~introduced
a hybrid sandpile/extremal dynamics model of seismicity
that also display waiting time 
distributions with power-law tails~\cite{lippiello05:_SOC_memory}.

In summary, 
we have studied a sandpile cellular automaton exhibiting self-organized
criticality
(even without open boundaries). A critical stationary state with 
correlated avalanches and intermittent transport 
takes place because the model is perturbed at
a position slowly diffusing in space,
a process leading to clustered nucleation points and 
 motivated by the observation of natural phenomena like earthquakes, 
where epicenters are correlated.  
Indeed,
the distributions of waiting times between avalanches larger 
than given thresholds  have striking and unique
similarities with the ones found in real seismicity.
We can thus reproduce some of 
the complexity of natural critical phenomena,
involving jointly energetic, spatial and temporal aspects.
Our results suggest that the logic that led to the present model is right,
namely that spatial correlations should not be arbitrarily removed by imposing
a random drive in models of critical phenomena like
sandpiles, which are by themselves already very simplified objects.
The positive comparison of our results with earthquake activity leads to
propose that a rugged landscape as depicted  in Fig.~\ref{fig:1}(a) may
be related to the stress distribution along a fault.
The absence of a single slope profile (as observed
in sandpiles with open boundaries) implies that
system-wide events are not always possible,
with potential interesting implications about predictability of large events.

\acknowledgments 
We thank S.~Boettcher, 
M.~De~Menech, M.~Paczuski, A.~L.~Stella and C.~Vanderzande for
useful discussions. M.~B.~acknowledges support from a FWO
 position (Flanders).


\begin{thebibliography}{10}
\bibitem{note0}
We chose not to use sequential updates because the model is not Abelian.
\bibitem{note_BC}
With periodic BC, no grains escape from the sandpile, and its
average height increases as $t/L$.
\bibitem{note1}
The etsimate of $D$ is easily obtained by matching all the cutoffs of the
distributions on a single value, while the value $\tau=1$ is
deduced by the good data collapse and by the fact that a direct
fit of the slopes of the distribution on Fig.~\ref{fig:Pa} would
give $\tau\approx 0.9$ while an analysis of the momenta $\langle
a^q \rangle$ yields $\tau \gtrsim 1$. Hence, a consistent
conclusion is that $\tau\simeq1 $ and that its direct estimates
are slightly contrasting due to finite size corrections to
scaling.
\bibitem{note_Rs}
$R(s)$ is the total number of events with size
$\ge s$ divided by the total number of added grains.
\end{thebibliography}
\end{document}